\documentclass{iauc}
\usepackage{graphicx}
\pubyear{2005}
\volume{199}
\pagerange{1--}
\jname{Probing Galaxies through Quasar Absorption Lines}
\editors{P. R. Williams, C. Shu, and B. M\'{e}nard, eds.}
\title[Highly ionized absorbers] %% give here short title %% 
{Highly ionized absorbers at high redshift}

\author[Bergeron \& Herbert-Fort]   %% give here short author list %%
{Jacqueline Bergeron$^1$  
\and St\'ephane Herbert-Fort$^1$} 
\affiliation{$^1$Institut d'Astrophysique de Paris-CNRS, 98bis Boulevard Arago,
F-75014, Paris, France \break 
email: bergeron@iap.fr
}

\begin{document}

\maketitle

\begin{abstract}
We build a sample of O\,{\sc vi} absorption sytems in the redshift 
range 2.0 $\lesssim z \lesssim$ 2.6 using high spectral 
resolution data of ten  quasars from the {\it VLT-UVES} Large Programme. 
We investigate the existence of a metal-rich O\,{\sc vi} population 
and define observational criteria for this class of aborbers under the 
assumption of photoionization. The low temperatures of nearly half of 
all O\,{\sc vi} aborbers, implied by their line widths, are too low for 
collisional ionization to be a dominant process. We  estimate  
the oxygen abundance  under the assumption of photoionization; a striking 
result is the bimodal distribution of [O/H] with median values close to 
0.01 and 0.5 solar for the metal-poor and metal-rich populations, 
respectively. Using the line widths to fix the 
temperature or assuming a constant, low gas density does not drastically
change  the metallicities of the metal-rich population. We present the 
first estimate of the O\,{\sc vi} column density distribution. Assuming a single 
power-law distribution, $f$(N) $\propto$ N$^{-\alpha}$,  yields 
$\alpha \sim 1.7$ and a normalization 
of $f$(N) $ = 2.3\times 10^{-13}$  at log N(O\,{\sc vi}) $\sim$ 13.5, both 
with a $\sim$30\% uncertainty. The value of $\alpha$ is similar to that  
 found for C\,{\sc iv} surveys, whereas the normalization factor is 
about ten times higher. 
We use $f$(N) to derive the number density 
per unit $z$ and cosmic density, $\Omega_{\rm b}$(O\,{\sc vi}), selecting 
a limited column density range not strongly affected by incompleteness 
or sample variance. Comparing our results with those obtained at $z\sim0.1$ 
for a similar range of column densities implies some decline of $dn/dz$ 
with $z$. The cosmic O\,{\sc vi} density derived from $f$(N), 
$\Omega_{\rm b}$(O\,{\sc vi})$\approx (3.5\pm ^{3.2}_{0.9}) \times 10^{-7}$, 
 is 2.3 times higher than the value estimated using the observed O\,{\sc vi} 
sample (of which the metal-rich population contributes $\sim$35\%),  
easing the problem of missing metals at high $z$ ($\sim$ 1/4 of the produced 
metals) but not solving it. We find that the 
majority of the metal-rich absorbers are located within $\sim$ 450 km s$^{-1}$
of strong Ly-$\alpha$ lines and show that, contrary to the metal-poor absorbers, 
 this population cannot be in hydrostatic equilibrium. All of the O\,{\sc vi} 
absorber properties imply that there are two distinct populations:  
metal-poor absorbers tracing the intergalactic medium and  metal-rich absorbers 
 associated with active sites of star formation and most probably linked to  
galactic winds.

%% add here a maximum of 10 keywords, to be taken form the file <Keywords.txt>.

\end{abstract}

\firstsection % if your document starts with a section,
\section{Introduction}

There are two major open questions that could be solved by the existence 
of a warm-hot and/or highly ionized phase of the intergalactic medium 
(IGM): the missing baryons at low redshift, $z \sim 0$-0.5, and the missing
metals at $z \sim 2.5$. 

The baryon budget at low $z$ implies that about 45\% of the cosmic baryons are 
still in the form of ionized gas in the IGM (\cite 
%(\cite [Fukugita, Hogan \& Peebles (1998)]
{Fukugita98}). The missing baryonic matter could reside in a warm-hot IGM 
(WHIM) as predicted by hierarchical structure formation models (see e.g. 
\cite [Cen \& Ostriker 1999]{Cen99}; \cite {Dave01}). The cooler phase of the 
WHIM can be probed by O\,{\sc vi}$\lambda\lambda1031,1037$ absorption but, at 
the sensitivity of the {\it FUSE} and {\it HST} surveys, the contribution to the 
cosmic baryon density, $\Omega_{\rm b}$, of the detected O\,{\sc vi} absorbers 
is only $\sim$5\% (\cite {savage02}; \cite {richter04}).
The hotter phase of the WHIM, $T>5\times10^5$ K, can be probed by 
O\,{\sc vii}-O\,{\sc viii} X-ray absorption. There are very few suitably bright
targets for X-ray spectroscopy with {\it Chandra} and {\it XMM-Newton},   
 and the rare confirmed detections, at a significance level higher 
than 3$\sigma$ (Nicastro \etal\ 2005a,b), have O\,{\sc vii} column densities 
about ten times larger than those of O\,{\sc vi}.  The contribution of this 
intergalactic hot phase to  $\Omega_{\rm b}$ could be up to ten times 
higher than that of the cooler WHIM.
%(\cite {nicastro05a}; \cite {nicastro05b})

At $z \sim 2.5$, at least 90\% of the baryons are in the Ly-$\alpha$ forest,
but only about 10\% of the metals produced by star formation activity in 
Lyman Break Galaxies (LBGs) have been detected up to now (\cite {pett99}).
%The amount of metals produced by LBGs enrich the cosmic baryons to about 
%1/25 of solar metallicity. Using more recent estimates gives a somewhat
The mean metal enrichment of the IGM could reach  a value 
Z $\simeq$ 0.04 Z$_{\odot}$ (\cite {pett99}) and recent simulations of 
galactic winds give estimates in the range 0.01-0.06 Z$_{\odot}$ 
(\cite {bert05}). The observed C\,{\sc iv} cosmic density equals  
$\Omega_{\rm b}$(C\,{\sc iv}) $\sim 7\times10^{-8}$ (\cite {song01}; 
\cite {scann05}) and, assuming an ionization correction of about a factor two,
leads to a cosmic abundance [C/H]$\sim-2.9$, thus a shortfall of metals by
a factor of at least ten and maybe up to $10^2$. The missing metals could reside 
in hot gaseous halos around star-forming galaxies (\cite {pett99};
\cite {ferr05}) and the 
cooler part of these hot bubbles might be traced by O\,{\sc vi} absorption. 

A few surveys of O\,{\sc vi} absorbers at $z \sim 2.0$-2.5 have already been 
conducted at the {\it VLT} and {\it Keck} telescopes, some for only a limited 
number of sightlines (\cite {berg02}; \cite {car02}; \cite {sim02},2004). A 
non-negligible fraction, $\sim$1/3, of the O\,{\sc vi} absorptions associated
with the Ly-$\alpha$ forest have line widths $b<14$ km s$^{-1}$, thus 
$T<2\times10^5$ K, which favors a radiative ionization process. A hard UV 
background flux, i.e. small discontinuity at 4 Ryd (\cite {haa96}), reproduces 
well the observed ionic ratios for $-3.0<$ [Z/H] $<-0.5$. The inferred values of 
$\Omega_{\rm b}$(O\,{\sc vi}) of the above surveys are $\approx 1.1 
\times10^{-7}$  (assuming $\Omega_{\rm \Lambda}, \Omega_{\rm m}, \Omega_{\rm b}, 
h=$ 0.7, 0.3, 
0.04, 70 throughout this paper). Applying a conservative ionization correction, 
O\,{\sc vi}/O=0.15, yields a mean oxygen abundance of [O/H]$\sim-2.7$, and thus,
as for  C\,{\sc iv} surveys, leaves open the problem of the missing metals.

However, a higher metallicity has been derived for  O\,{\sc v} absorbers. 
The EUV O\,{\sc v}$\lambda630$ singlet was searched for and detected in a 
stacked composite absorption spectra from {\it HST-FOS} data for absorbers 
at $1.6<z<2.9$ with 
a large range of H\,{\sc i} column densities (\cite {telf02}). Except in the 
strongest H\,{\sc i} systems, the lack of detection of the associated EUV 
O\,{\sc iv} doublet also suggests a hard ionizing background flux, and the 
derived oxygen abundance is [O/H]$\sim-2.2$ to $-1.3$.

The paper is organized as follows: our new {\it VLT} O\,{\sc vi} sample is 
presented in \S\ 2. In \S\ 3, we give results on the oxygen abundance 
derived under various assumptions on the ionization process. The O\,{\sc vi} 
column density distribution and the contribution of the O\,{\sc vi} absorbers 
to the cosmic baryon  density are given in \S\ 4. The origin of these 
absorbers is discussed in \S\ 5.
We present our conclusions and prospectives in \S\ 6.

\section{The O\,{\sc vi} sample}

The {\it VLT-UVES} Large Programme ``The Cosmic Evolution of the IGM'' provides
a homogeneous sample of quasar sightlines, with emphasis given to lower 
redshift quasars ($z<3$) to take advantage of the high UV sensitivity of 
{\it UVES}.
This allows a study of O\,{\sc vi} systems in the range $z=2.0$-2.5 where the 
crowding of the Ly-$\alpha$ forest is not too severe. Altogether, the sample
comprises 21 bright quasars (most with V $<$ 17), of which 19 are at $2<z<4$,
observed with dichroics blue and red. The spectral resolution is $R =45,000$ 
(line width  $b$ = 6.6 km s$^{-1}$) and the exposure time per setting per 
quasar (2 settings per quasar) of 6 to 10 hr yields a signal-to-noise 
S/N $\sim$ 30-40 and 100 at 3300 and 5500 \AA\ respectively.
The data were reduced using an upgraded version of the {\it ESO-UVES} 
data-reduction pipeline (Aracil  \etal\, in preparation).

We present results derived from the spectral analysis of ten quasars at  
2.1 $<z_{em}<$ 2.8. Our O\,{\sc vi} sample comprises 136 absorbers with column 
densities in the range 12.7 $<$ log N(O\,{\sc vi}) $<$ 14.6. These absorbers span 
the redshift interval 1.99 $<z<$ 2.57 with a mean value of $\overline{z}=2.28$.
Due to partial blending of the associated H\,{\sc i} absorptions 
and small velocity differences between the O\,{\sc vi} and H\,{\sc i} components,
we group the O\,{\sc vi} and H\,{\sc i} absorptions into 51 systems.  
Absorption systems within 5000 km s$^{-1}$ of the quasar 
emission redshift are excluded from this sample.

%%%%%%%%%%%%%%%%%%%%%%%%%%%%%%%%%%%%%%%%%%%%%%%%%%%%%%%%%%%%%%%%%%%%%%%%%%

\subsection{The O\,{\sc vi} subsamples}

There are unusual O\,{\sc vi} absorbers with high abundances, 
$-1<$ [O/H] $\lesssim 0$, in 
%[O/H] $>-1$ (or even [O/H] $\sim$ 0), in 
%two of the three 
previous O\,{\sc vi} surveys
(\cite {berg02}; \cite {car02}). They have high ionic ratios, 
N(O\,{\sc vi})/N(H\,{\sc i}) $>$ 0.5, and low H\,{\sc i} column densities,
log N(H\,{\sc i}) $<$ 13.0. The  survey by Simcoe \etal\ (2004) focussed
on systems with log N(H\,{\sc i}) $>$ 13.6, which could account for the 
underrepresentation of highly  metal-rich O\,{\sc vi} absorbers in their
O\,{\sc vi} sample.
Since these intriguing systems are not present in every sightline, a large
quasar sample is  mandatory for a statistically significant number 
of metal-rich O\,{\sc vi} absorbers.

It is possible to define the class of metal-rich absorbers using observed
column density ratios derived from photoionization models since the small
line widths of a non-negligible fraction of the O\,{\sc vi} systems imply 
``low'' gas temperatures (see \S\ 3.1). Adopting a hard UV background 
spectrum together with a 0.1 solar metallicity leads to observational 
identification criteria for the following classes of absorbers:\\
- type 1: N(O\,{\sc vi})/N(H\,{\sc i}) $>$ 0.25: metal-rich absorbers,\\
- type 0: N(O\,{\sc vi})/N(H\,{\sc i}) $<$ 0.25: metal-poor absorbers.

\begin{figure}[!hb]
\center
%\hspace{-3.3cm}
  \includegraphics[width=10.2cm,angle=90]{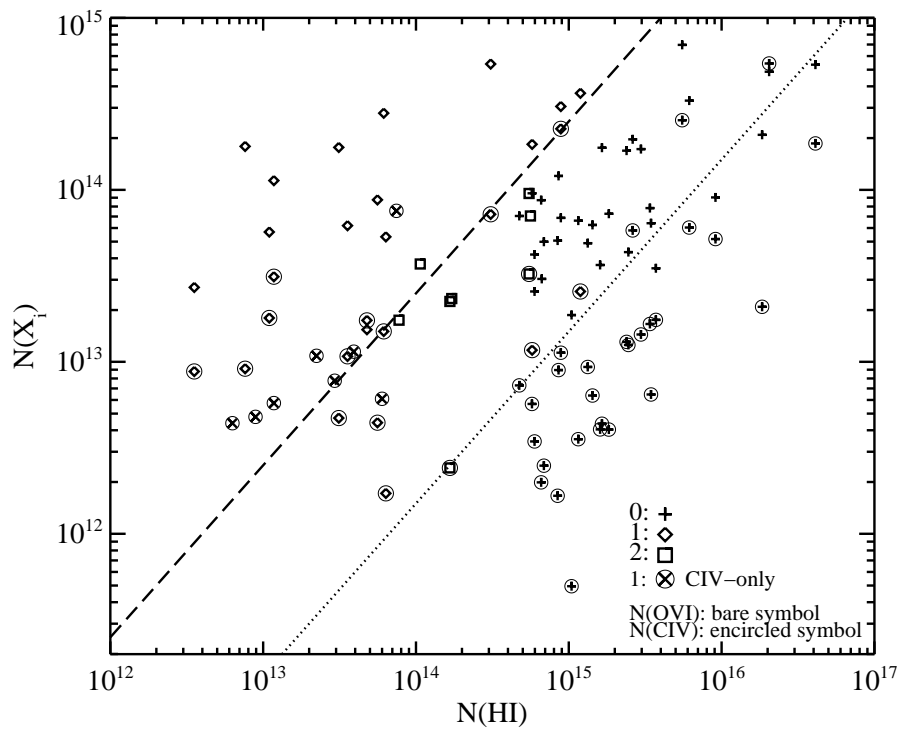}
%\vspace{-1.3cm}
  \caption{Metal column densities of O\,{\sc vi} and C\,{\sc iv} versus
           H\,{\sc i} column density. The dashed and dotted line give the 
locations of systems with N(O\,{\sc vi})/N(H\,{\sc i}) = 0.25  and  
N(C\,{\sc iv})/N(H\,{\sc i}) = 0.015, respectively.}
\label{fig:OCvsH}
\end{figure}

\begin{figure}[!ht]
\center
\includegraphics[width=11cm,angle=0]{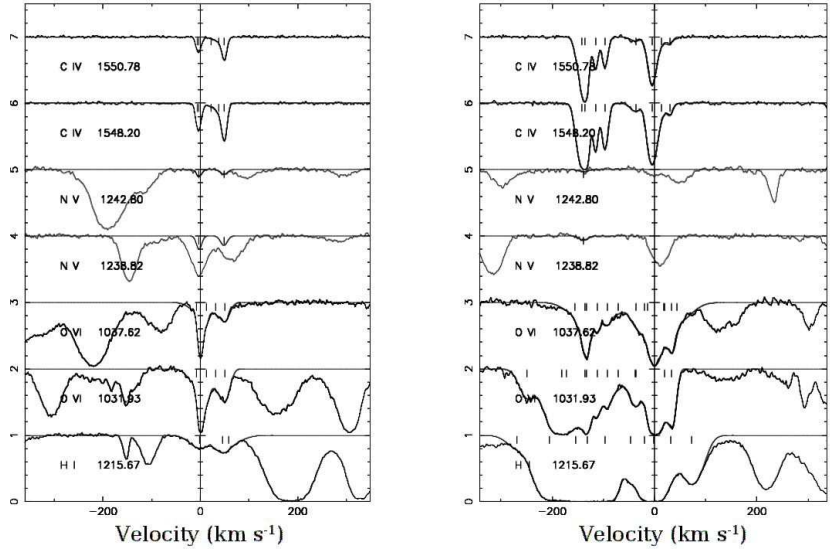}
\caption{H\,{\sc i} and metal absorptions of type 1 O\,{\sc vi} absorbers: a 
weak H\,{\sc i} system at $z=2.468$ is shown in the left panel, and a strong 
H\,{\sc i} system at $z=2.398$ is shown in the right panel. The latter
is 140 km s$^{-1}$ away from a type 0 absorber. }
\label{fig:type1}
\end{figure}

\hspace{.4cm}

\begin{figure}[!h]
\center
\includegraphics[width=11cm,angle=0]{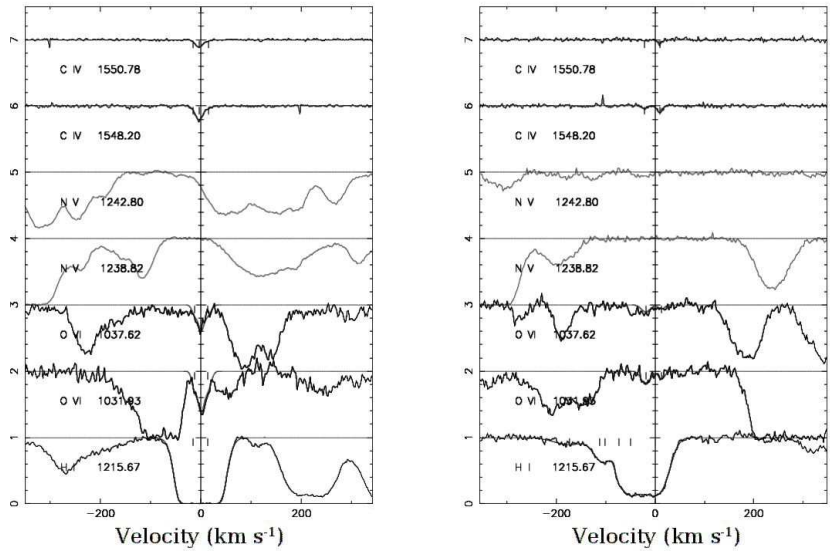}
\caption{H\,{\sc i} and metal absorptions of a type 0 O\,{\sc vi}
absorber at $z=2.089$ (left panel) and a type 2 O\,{\sc vi} absorber 
at $z=2.314$ (right panel).}
\label{fig:type0}
\end{figure}

%\hspace{.7cm}

\begin{figure}[!ht]
\vspace{.7cm}
\center
\includegraphics[width=11cm,angle=0]{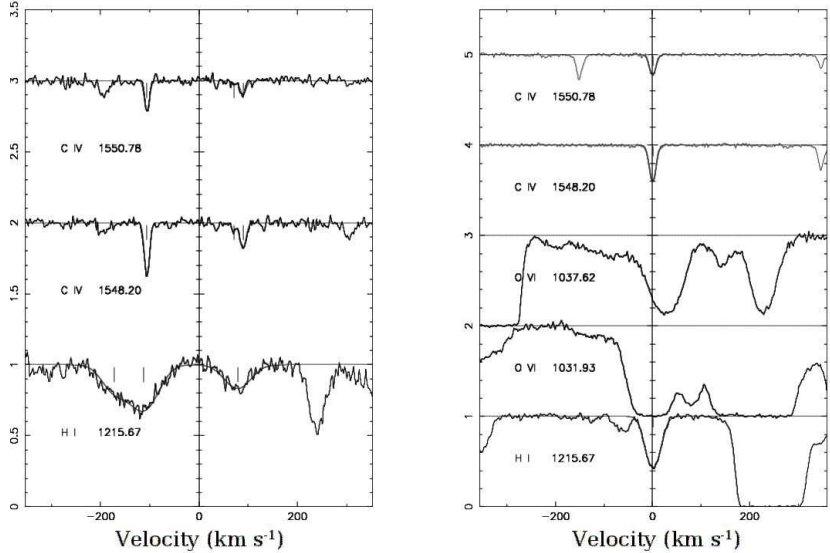}
%\hspace{2cm}
%\vspace{.7cm}
  \caption{H\,{\sc i} and metal absorptions of C\,{\sc iv}-only  type 1  
absorbers: two low redshift systems at $z=1.727$ and 1.729 (left panel) and
one system at $z=2.415$ with O\,{\sc vi} lines fully blended with saturated
Lyman lines (right panel).}
\label{fig:typeciv}
\end{figure}

\hspace{.7cm}

\noindent
There are 39 O\,{\sc vi} type 1 components, 12.9 $<$ log N(O\,{\sc vi}) 
$<$ 14.5, grouped in 14  O\,{\sc vi}-H\,{\sc i} systems.

A similar criterium is derived for the C\,{\sc iv} systems and is used 
to identify C\,{\sc iv}-only metal-rich absorbers (O\,{\sc vi} doublet
either outside the observing range, $z<2.0$, or fully blended with saturated 
Lyman lines):\\
- C\,{\sc iv}-only type 1: N(C\,{\sc iv})/N(H\,{\sc i}) $>$ 0.015: metal-rich 
absorbers.\\
There are 18 C\,{\sc iv}-only type 1 components, 11.8 $<$ log N(C\,{\sc iv}) 
$<$ 13.8, grouped in 8  C\,{\sc iv}-H\,{\sc i} systems.

Finally, a few absorbers with O\,{\sc vi} blended with strong Lyman lines,
and thus with uncertain values of N(O\,{\sc vi}), are labelled type 2.

These different classes of absorbers are shown  in figure~\ref{fig:OCvsH}.
About 70\% of the O\,{\sc vi}+C\,{\sc iv}-only type 1 absorbers have 
weak associated H\,{\sc i} lines, log N(H\,{\sc i}) $<$ 13.6, and 
the type 0 and type 1 O\,{\sc vi} absorbers span roughly the same 
N(O\,{\sc vi}) range. This demonstrates the importance of searching for
O\,{\sc vi} systems whatever the strength of their associated H\,{\sc i} 
aborption.

Examples of type 1 absorbers are given in figure~\ref{fig:type1}. The 
proximity in velocity space of a strong Ly-$\alpha$ system to a 
type 1 absorber is investigated in \S\ 5.1.
  Examples of types 0 \& 2 O\,{\sc vi} absorbers and C\,{\sc iv}-only  type 1  
absorbers are  given in figures \ref{fig:type0} and 
  \ref{fig:typeciv}, respectively.

%%%%%%%%%%%%%%%%%%%%%%%%%%%%%%%%%%%%%%%%%%%%%%%%%%%%%%%%%%%%%%%%%%%%%%%%%%%%

\section{Abundances}

\subsection{O\,{\sc vi} line widths} 

\begin{figure}[!ht]
\center
\hspace{-.0cm}
  \includegraphics[width=7cm,angle=90]{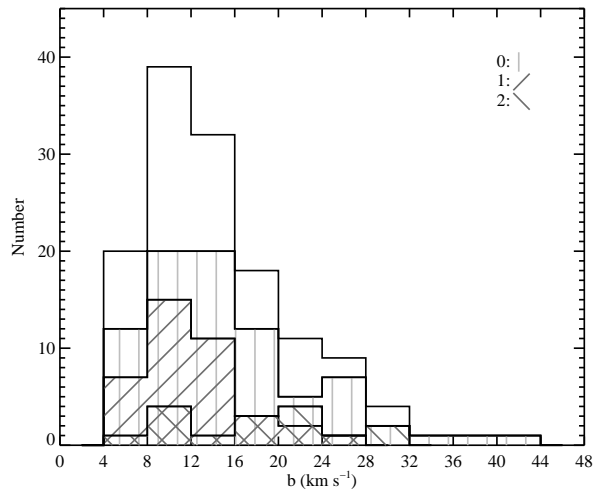}
\vspace{-0.1cm}
  \caption{Distribution of the O\,{\sc vi} line widths
with the totals outlined.}
\label{fig:histb}
\end{figure}

The histogram of O\,{\sc vi} line widths is shown in figure~\ref{fig:histb}.
There are 81, 39 and 16 O\,{\sc vi} components for the types 0, 1 and 2 
absorbers, respectively. The bulk of the $b$ distributions of the types 0
and 1 overlap, but not their high velocity tails. A  Kolmogorov-Smirnov test 
shows that these distributions indeed  differ at the 98\% confidence level. 
It should be stressed that among the broader absorbers, $b>16$ km s$^{-1}$, 
most components are blends of several O\,{\sc vi} lines within a velocity 
range of a few tens of km s$^{-1}$. Very few individual 
components are unambiguously broad. 

Close to half (43\%) of the O\,{\sc vi} absorbers have line widths 
$b<12$ km s$^{-1}$. This confirms the results previously found with smaller 
O\,{\sc vi} samples. The implied temperatures, $T <1.4\times 10^5$ K, as also 
found for the non-saturated associated Lyman lines, are too low for O\,{\sc vi}
to be produced by collisional ionization even for abundances close to 
solar.  We will thus assume that photoionization is the dominant ionization 
process, but will also consider simple cases where there is additional 
collisional heating of the gas, possibly through shocks.

\vspace{-0.2cm}

\subsection{Abundances under the assumption of photoionization}

Following previous discussions on constraints of the spectral energy distribution 
of the ionizing background flux (e.g. \cite {berg02}; \cite {car02};
 \cite {telf02}), we select a hard UV metagalactic flux (\cite {haa96}) to 
derive the gas ionization level. We used the CLOUDY v94.0 code (\cite {ferl98})
to estimate ionic column density ratios as a function of the ionization 
parameter, $U \equiv n_{\rm \nu}/n_{\rm H}$, and assumed solar relative 
abundances (\cite {and89}).   
For each system, the value of $U$ is fixed by the observed ionic ratio 
N(O\,{\sc vi})/N(C\,{\sc iv}) which is applicable only if O\,{\sc vi} and 
C\,{\sc iv} are in the same phase. This should be the case for most absorbers
as Si\,{\sc iv} is not detected, except in a few systems with large 
N(H\,{\sc i}) ($>10^{15}$ cm $^{-2}$). The observed range of 
N(O\,{\sc vi})/N(C\,{\sc iv}) implies $-1.4 \leq $ log $U \leq -0.4$, 
thus an ionization ratio  0.09~$\leq $~O\,{\sc vi}/O~$\leq $~0.21.

\begin{figure}[!ht]
\center
  \includegraphics[width=7cm,angle=90]{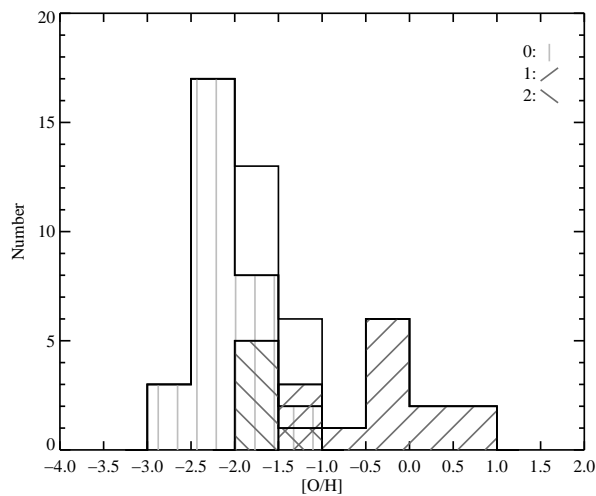}
  \caption{Distribution of the oxygen abundance in the photoionization case.
The [O/H] median values for the types 0, 1 and 2 populations are 
$-$2.07, $-$0.33 and $-$1.56, respectively.}
\label{fig:histO_H}
\end{figure}

The distributions of the derived oxygen abundances are presented 
in figure~\ref{fig:histO_H} for the 31, 14 and 6 O\,{\sc vi} systems 
 of type 0, 1 and 2, respectively.
Contrary to the distribution of $b$(O\,{\sc vi}) shown above,
there is very little overlap between the [O/H] distributions of the types 0 
and 1 populations. This further suggests that they are indeed two distinct 
populations.  The type 2 [O/H] distribution spans a small range in between 
those of the other two populations.

\begin{figure}[!h]
\center
\vspace{0.5cm}
\hspace{.8cm}
  \includegraphics[width=6cm,angle=0]{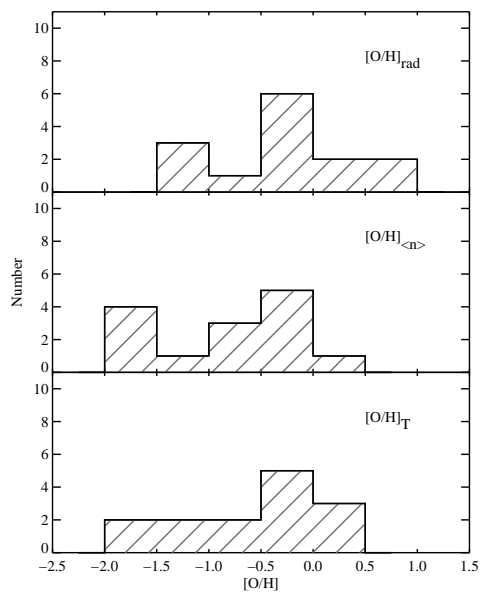}
\vspace{.7cm}
  \caption{Distributions of the oxygen abundance for the type 1 population 
under various ionization conditions: (1) pure photoionization case
(top panel), (2) constant gas density (middle panel), 
(3) photoionization plus temparature fixed by $b$(O\,{\sc vi})
(bottom panel). 
The [O/H] median values for the cases 1, 2 and 3  are 
$-$0.33, $-$0.80 and $-$0.35, respectively.}
\label{fig:histO_Hall}
\end{figure}

\subsection{Abundances under the assumption of other ionization processes}

To confirm the difference in metallicity between the type 0  (population
tracing the IGM) and type 1 (population tracing highly metal-enriched sites)
absorbers, we investigate 
%other ionization conditions for the type 1 population. 
whether other ionization conditions could yield much lower abundances 
for the type 1 population. 

First, we consider a gas phase of constant density, $\rho$, thus a single 
value of the ionization paramater. We select an overdensity 
$\delta\equiv(\rho/\overline{\rho}) \approx 10$ at $z \approx 2.2$ which is 
within the range of values found in the previous O\,{\sc vi} surveys.
This yields log $U=-0.5$, thus an ionization ratio  O\,{\sc vi}/O = 0.16. 
In a large fraction of these cases, O\,{\sc vi} and 
C\,{\sc iv} do not trace the same phase.

Secondly, we reconsider photoionization by a hard UV background flux  but now 
with the gas temperature derived from the $b$ value of the main O\,{\sc vi} 
component of each system. This is 
to account for possible additional shock heating.
The value of $U$ is still derived from the observed N(O\,{\sc vi})/N(C\,{\sc iv})
 ionic ratio. There is no solution for absorbers with $T >2.0\times 10^5$ K, or 
$b>14$ km s$^{-1}$, implying that at these higher temperatures O\,{\sc vi} and 
C\,{\sc iv} are not co-spatial. However, for the type 1 systems, the main 
O\,{\sc vi} component is always narrower than 14 km s$^{-1}$ except in the  
case of one O\,{\sc vi} doublet blended  with Lyman lines.
 
Together with the case of pure photoionization, the distributions of the 
oxygen abundances estimated in the above two cases  are shown in 
figure~\ref{fig:histO_Hall}. Although the values of [O/H] are somewhat lower 
under the new ionization conditions, they remain far higher than those of 
the type 0 population. This confirms that the types 0 and 1 O\,{\sc vi} 
absorbers trace markedly different populations.

%%%%%%%%%%%%%%%%%%%%%%%%%%%%%%%%%%%%%%%%%%%%%%%%%%%%%%%%%%%%%%%%%%%%%%%%

\section{Contribution of the O\,{\sc vi} absorbers to the cosmic baryon 
 density}

\subsection{Column density distribution}

The column density distribution, $f$(N), of O\,{\sc vi} absorbers per 
unit redshift path per unit column density can be written:
\begin{equation}
f{\rm (N)} =  \{ n/(\Delta {\rm N} \sum \Delta X) \}, 
 \label{f}
\end{equation}
where $n$ is the number of O\,{\sc vi} absorbers in a column density bin 
$\Delta {\rm N}$ centered on N for a sample of quasars with total redshift 
path $\sum \Delta X$. For our adopted cosmology, the redshift path is 
defined as:
\begin{equation}
 {\rm d}X \equiv (1+z)^2 \{\Omega_{\rm \Lambda}+\Omega_{\rm m}(1+z)^3\}^{-0.5}
  \  {\rm d}z ,
 \label{X}
\end{equation}
\[
{\rm or} \ {\rm d}X/{\rm d}z \cong \{(1+z)/0.3\}^{0.5} \ {\rm when} \  z>1.
\]

\begin{figure}[!ht]
\center
%\hspace{-.0cm}
  \includegraphics[width=8cm,angle=90]{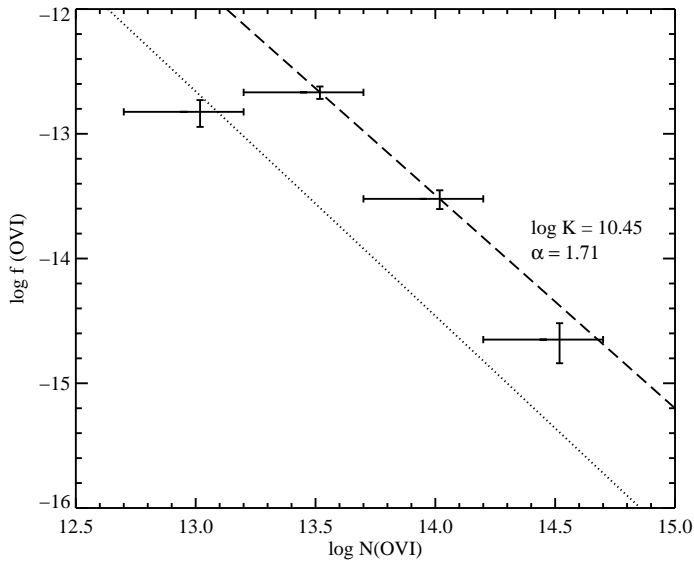}
%\vspace{-1.0cm}
  \caption{Column density distribution of O\,{\sc vi} absorbers.
The dashed line is the fit to our data in the column density range
13.0 $<$ log N(O\,{\sc vi}) $<$ 14.3 (see text). 
The dotted line is the fit to the  column density distribution 
of C\,{\sc iv} absorbers given by Songaila (2001).}
\label{fig:Ndistr}
\end{figure}

The O\,{\sc vi} column density distribution is shown in figure~\ref{fig:Ndistr}.
It can be seen that the present data become incomplete below a column density
of $\sim 1\times 10^{13}$ cm$^{-2}$ and that sample variance may be important 
at column densities larger than $\sim 2\times 10^{14}$ cm$^{-2}$. In between 
these limits, a power-law fit ($f$(N) $\propto$ N$^{-\alpha}$)
gives  $\alpha \simeq 1.7$. To estimate the uncertainty in $\alpha$, we shift 
the $\Delta$N bins by 0.1 dex 
and derive new power-law fits. This yields $\alpha = 1.71 \pm ^{0.48}_{0.47}$  
and a normalization of $f$(N) $ = 2.3\times 10^{-13}$  at log 
N(O\,{\sc vi}) = 13.5, with a $\sim$30\% uncertainty.

The value of the power-law index is similar to that obtained from  C\,{\sc iv} 
samples, $\alpha$(C\,{\sc iv}) $\simeq 1.8$, such as those drawn from 
the {\it VLT-UVES} Large 
Programme at $\overline{z}$(C\,{\sc iv}) = 2.16 (\cite {scann05}) or from 
{\it Keck-HIRES} data at higher redshift (\cite {song01}). The power-law 
fit of the latter (corrected for the different adopted cosmological 
parameters) is also shown in figure~\ref{fig:Ndistr}. At 
N(O\,{\sc vi}) = N(C\,{\sc iv}) = $10^{13.5}$ cm$^{-2}$, the value of 
 $f$(N) for O\,{\sc vi} absorbers is nearly a factor of ten larger than that for  
C\,{\sc iv} absorbers. 

\subsection{Number density}

We use the power-law fit to $f$(N) for the O\,{\sc vi} population  to estimate 
the  number density per unit $z$ of O\,{\sc vi} absorbers: 
\begin{equation}
dn/dz = (dX/dz) \int f({\rm N) dN}.
\end{equation} 
We select conservative N(O\,{\sc vi}) limits of $10^{13}$ and $10^{15}$ cm$^{-2}$,  a range that is not drastically affected by incompleteness or sample variance. 
Adopting the fit with $\alpha$ = 1.71, we then get $dn/dz = 74$ at 
$\overline{z}=2.3$. Taking into account the range of possible values of 
the power-law index and normalization factor of $f$(N), we obtain
$66 < dn/dz < 106$.   

At low redshift, $\overline{z}=0.1$, surveys with {\it FUSE} and  {\it HST} 
give $dn/dz \approx 13 $ for a rest-equivalent width limit of 
$w_{\rm r, min}=50$~m\AA, or log N(O\,{\sc vi})=13.60
in the optically thin case (see  \cite {sem04}).  For this column density 
limit, we get $dn/dz \approx 26$ at $\overline{z}=2.3$, whereas we expect 
a somewhat higher value, $dn/dz \approx 36$, in the case of an 
unevolving O\,{\sc vi} population. 
However, comparison between the values of $dn/dz$ at $\overline{z}=0.1$
 and 2.3 is not straightforward as O\,{\sc vi} absorbers may trace  
different populations at low and high redshift.

\subsection{Cosmic density of {\rm O}\,{\sc vi} absorbers}

The O\,{\sc vi} cosmic density can be expressed as a mass fraction 
relative to the critical density, $\rho_{crit}$. It can be estimated either
from the individual, observed O\,{\sc vi} column densities   
or using the power-law fit to $f$(N) of O\,{\sc vi} absorbers   
to correct for incompleteness.  

\subsubsection{Observed {\rm O}\,{\sc vi} cosmic density}

The mean cosmic density of a given ion can be expressed as:
\begin{equation}
\Omega_{\rm b, ion} = \{ H_0 m_{\rm ion}/c \rho_{crit} \} 
\{\sum {\rm N_{\rm ion}}/\sum  \Delta X \} = 
2.20\times 10^{-22} \{\sum {\rm N_{\rm ion}}/\sum \Delta X \},  
  \label{Omega-obs}
\end{equation}
where $H_0$ is the Hubble constant, $m_{\rm ion}$ and $\sum {\rm N_{\rm ion}}$
the atomic mass and the sum of the column densities of the given ion, 
respectively,  
and  $\sum \Delta X$ the total redshift path.
For our O\,{\sc vi} sample, we obtain 
$\Omega_{\rm b}$(O\,{\sc vi})$ = 1.51 \times 10^{-7}$, a value higher than 
previous estimates by a factor 1.3 (\cite {sim04}: sample restricted to 
O\,{\sc vi} systems with strong, associated H\,{\sc i} absorption [see \S\ 2.1]) 
and 1.8 (\cite {car02}: two sightlines, none with very strong O\,{\sc vi} 
absorbers). The  contribution of the O\,{\sc vi} type 1 population to  
$\Omega_{\rm b}$(O\,{\sc vi}) is 35\%.

\subsubsection{{\rm O}\,{\sc vi} cosmic density corrected for incompleteness}

The mean cosmic density of O\,{\sc vi} ions can also be written as:

\begin{equation}
\Omega_{\rm b} = 2.20 \times 10^{-22} \int {\rm N} f({\rm N) dN}
  \label{Omega-f}
\end{equation}
Using our fit with $\alpha$ = 1.71 and the same N(O\,{\sc vi}) limits 
as in \S\ 4.2, 
we get  $\Omega_{\rm b}$(O\,{\sc vi}) $\approx 3.5 \times 10^{-7}$, 
thus an incompleteness correction factor of 2.3 at $\overline{z}=2.3$.
The uncertainty in the power-law fit of $f$(N) leads to values in the range   
$2.6\times10^{-7} < \Omega_{\rm b}$(O\,{\sc vi}) $< 6.7\times10^{-7}$.

To estimate the mean cosmic density of oxygen, $\Omega_{\rm b}$(O), we
use the O\,{\sc vi} mean ionization level obtained in the pure photoionization 
case, $\langle$O\,{\sc vi}/O$\rangle$ = 0.15 (see \S\ 3.2). Under the other 
ionization conditions investigated in \S\ 3.3, this ratio is either 
similar or smaller.   
We then get $\Omega_{\rm b}$(O) $\approx (2.3\pm ^{2.1}_{0.6}) \times 10^{-6}$.

Using the solar oxygen abundance given by Anders \& Grevesse (1989), yields:
\\  
\centerline {log~($\Omega_{\rm b}$(O)/$\Omega_{\rm b}$(O)$_{\odot}) = -2.22$.} 
This result 
demands attention for the two following reasons: (1) the above value is
close to the median of [O/H] found for the O\,{\sc vi}  type 0 absorbers (IGM),
[O/H] = $-$2.07, but well below that of the  O\,{\sc vi}  type 1 absorbers 
(metal-enriched sites), [O/H] = $-$0.33 (see figures~\ref{fig:histO_H} 
and \ref{fig:histO_Hall}), and (2) it is smaller than the mean metal enrichment
of the IGM by star-forming galaxies at $z \sim 2.5$ (Z$^{\rm SF}$) 
by a factor of about 3.7 and 6.6 when adopting the values of 
Z$^{\rm SF} \approx$ 1/45 and 1/25 solar as given  by Ferrara \etal\ (2005) 
and Pettini (1999), respectively. Consequently, there is still a shortfall of 
observed metals as compared to those produced by LBGs, but about a factor three 
smaller than previously thought. 

Our sample contains very few cases of unambiguously broad O\,{\sc vi} 
doublets ($b>16$ km s$^{-1}$) which could trace hotter parts 
($T>2.5\times10^5$ K) of metal-rich sites. However, if most of the gas in these 
sites is at even higher temperatures, oxygen will then mainly be in the form of 
O\,{\sc vii} and O\,{\sc viii} ions and not detectable with present-day 
X-ray satellites.

%%%%%%%%%%%%%%%%%%%%%%%%%%%%%%%%%%%%%%%%%%%%%%%%%%%%%%%%%%%%%%%%%%%%%%%%%%%%%

\section{Origin of the O\,{\sc vi} absorbers}

%\subsection{Nearest neighbour in velocity space}
\subsection{Nearest strong {\rm H}\,{\sc i} absorption system}

 From a pixel analysis of {\it VLT-UVES} Large Programme quasar  spectra,
Aracil \etal\  (2004) found that weak O\,{\sc vi} absorption associated
with weak H\,{\sc i} absorption (0.2 $< \tau$(Ly-$\alpha$) $<$ 1 or 
12.9 $<$ log N(H\,{\sc i}) $<$ 13.6 for $b$(H\,{\sc i}) = 30 km s$^{-1}$)
is  predominantly detected in the vicinity ($\Delta v \leq$ 300 km s$^{-1}$) 
%(\cite {arac04})
of strong H\,{\sc i} absorption ($\tau$(Ly-$\alpha$) $>$ 4). 
These authors  suggested that the  O\,{\sc vi} absorption 
arising in regions spatially close to strong Ly-$\alpha$ absorption
may be part of  outflows from overdense regions.

\begin{figure}[!ht]
\center
%\hspace{-.0cm}
  \includegraphics[width=8cm,angle=90]{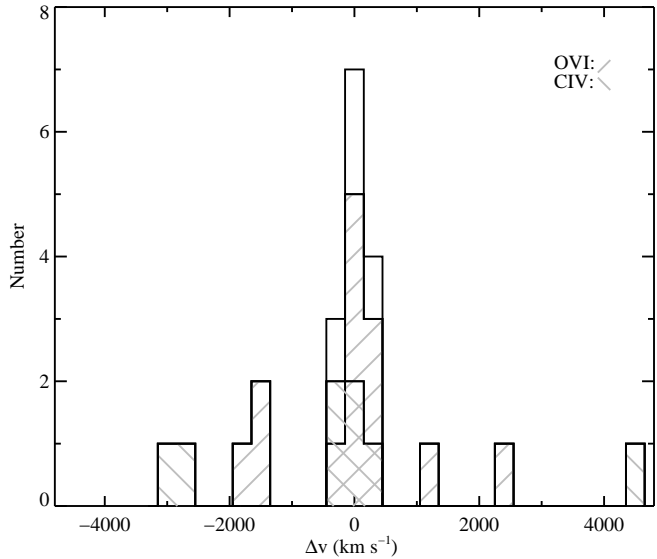}
%\vspace{-1.0cm}
  \caption{Distribution of the velocity difference between type 1 systems
and the nearest strong Ly-$\alpha$ system, with the totals outlined.}
\label{fig:dv}
\end{figure}

The O\,{\sc vi} type 1 population from our study should exhibit the same   
property as the  weak O\,{\sc vi} absorptions analyzed with the pixel analysis
method, since there is an overlap in their N(H\,{\sc i}) range. 
The distribution of $\Delta v$ between O\,{\sc vi} or   
C\,{\sc iv}-only type 1 systems and the nearest strong Ly-$\alpha$ system
($\tau$(Ly-$\alpha) > 4$) is presented in figure~\ref{fig:dv}. 
%$\tau$(Ly-$\alpha = 2.53$ N$_{14} \ b_{30}^{-1} 
For the few cases of an O\,{\sc vi} doublet associated with a saturated 
Ly-$\alpha$ line, $\Delta v$ is nul. It should be noted that 
all C\,{\sc iv}-only type 1 systems
have unsaturated, associated Ly-$\alpha$ lines. 

Among the O\,{\sc vi} and  C\,{\sc iv}-only type 1 systems,  64\% and 63\%,  
respectively, have a strong Ly-$\alpha$ system at $\Delta v <$ 450 km s$^{-1}$. 
Results from both the pixel analysis method of  weak O\,{\sc vi} 
systems (log $\tau$(O\,{\sc vi}) $\sim -1.35$) and the study of individual 
O\,{\sc vi} absorbers  suggest a  link to gas outflows.

\subsection{Gas density  of the O\,{\sc vi} absorbers}

The gas overdensity of the O\,{\sc vi} absorbers  is estimated for two cases: 
photoionization by a hard UV metagalactic flux and hydrostatic 
equilibrium (\cite {scha01}).

%$\delta\equiv(\rho/\overline{\rho})$,

\begin{figure}[!h]
\center
%\hspace{2.8cm}
  \includegraphics[width=6cm,angle=0]{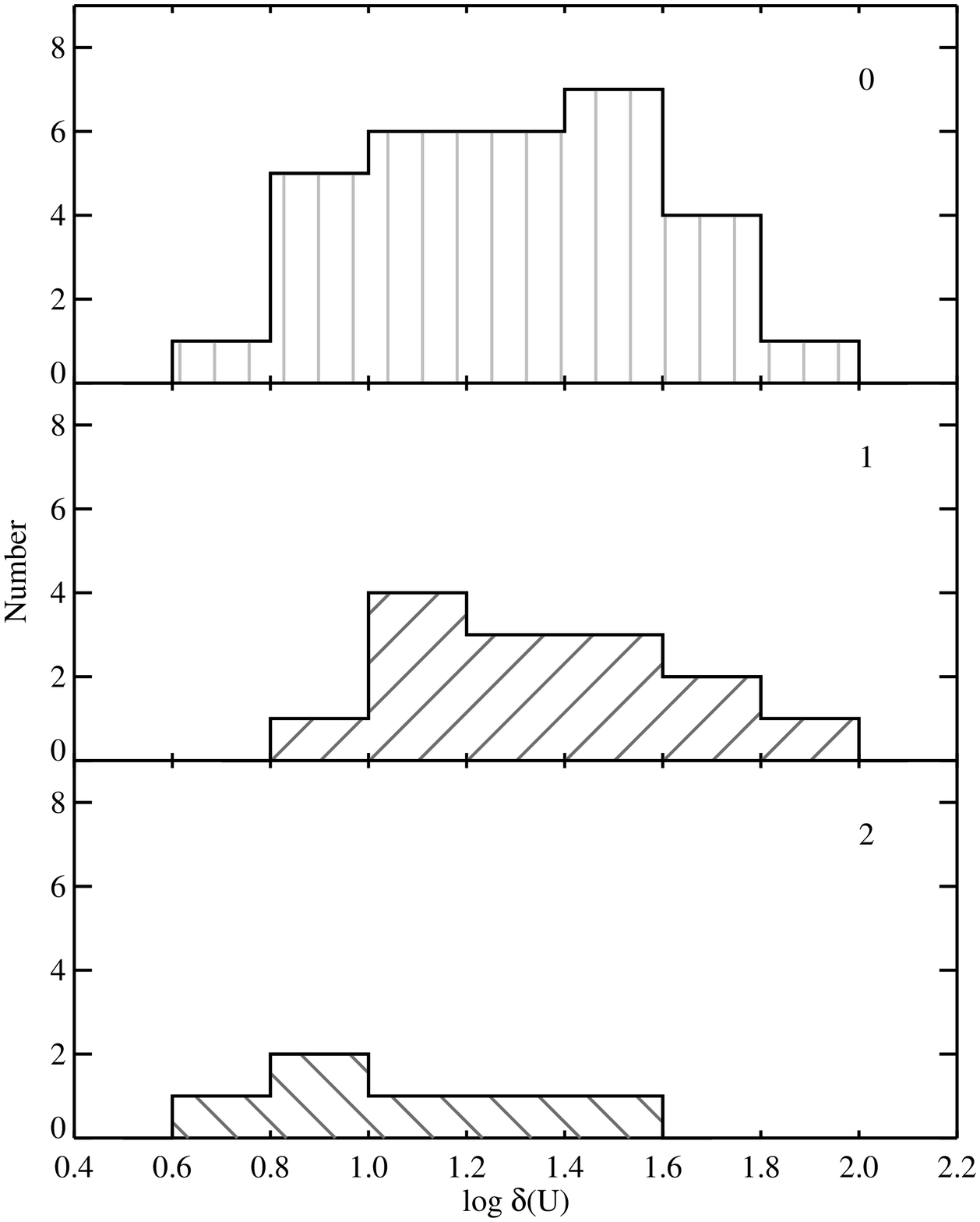}
\vspace{.7cm}
  \caption{Distribution of the O\,{\sc vi} absorber overdensity 
 in the photoionization case.}
\label{fig:rhoU}
\vspace{.7cm}
%\hspace{2.8cm}
\center
  \includegraphics[width=6cm,angle=0]{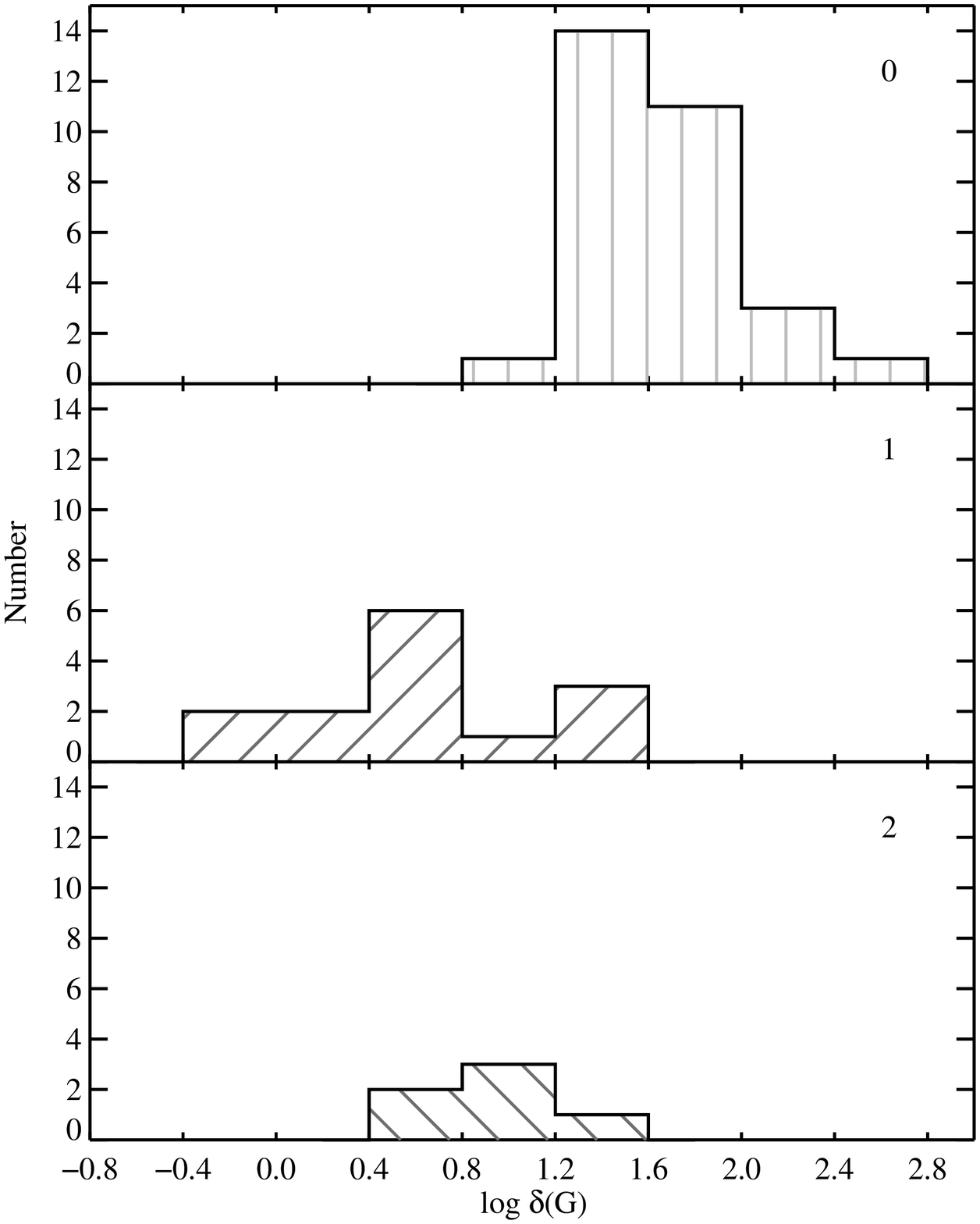}
\vspace{.7cm}
  \caption{Distribution of the O\,{\sc vi} absorber overdensity 
 in the hydrostatic equilibrium  case.}
\label{fig:rhoG}
\end{figure}

In the photoionization case $U$ is fixed by the O\,{\sc vi}/C\,{\sc iv} ionic 
ratio, assuming a relative O/C solar abundance. In the range $2.0<z<2.5$, 
the adopted hydrogen photoionization rate is 
$\Gamma$(H\,{\sc i}) $ \approx 1.5 \times 10^{-12}$ s$^{-1}$.
Using the  mean baryon density at each $z$(O\,{\sc vi}), we get:

\begin{equation}
\delta (U) \approx 4.0 \ U^{-1} ([1+z]/3)^{-3}.
\end{equation}
The results are shown in figure~\ref{fig:rhoU} for the different types of 
O\,{\sc vi} absorbers. The median values of $\delta(U)$ for the type 0 (metal-poor) 
and type 1  (metal-rich) populations are equal, $\delta (U) = 22$, and that of 
the type 2 population is $\sim$40\% smaller. A Kolmogorov-Smirnov test shows that 
the types 0 and 1 populations have the same $\delta(U)$ distribution 
at the 97\% confidence level. 

For the hydrostatic equilibrium case (\cite {scha01}: equation (8)), we assume 
a gas temperature $T = 4\times 10^4$ K and the same photoionization rate 
as above. This gives: 

\begin{equation}
\delta (G) = 3.7 \times 10 ^{-9} \ {\rm N(H\,{\sc {\i}})}^{2/3}([1+z]/3)^{-3}.
\end{equation}

 The results are presented in figure~\ref{fig:rhoG}. Contrary to the 
photoionization case, there is a marked difference 
between the types 0 and  1 populations. The median values of $\delta(G)$ are 
41 and 4.6 for the type 0 and type 1 absorbers, respectively, and that for the 
type 2 absorbers is 8.3. Moreover, the values of $\delta(G)$ for $\sim$80\% of 
the type 1 population (log $\delta(G)<1.1$) do not overlap with those obtained 
for the type 0 population.

\begin{figure}[!h]
\center
\hspace{-0.7cm}
  \includegraphics[width=9cm,angle=90]{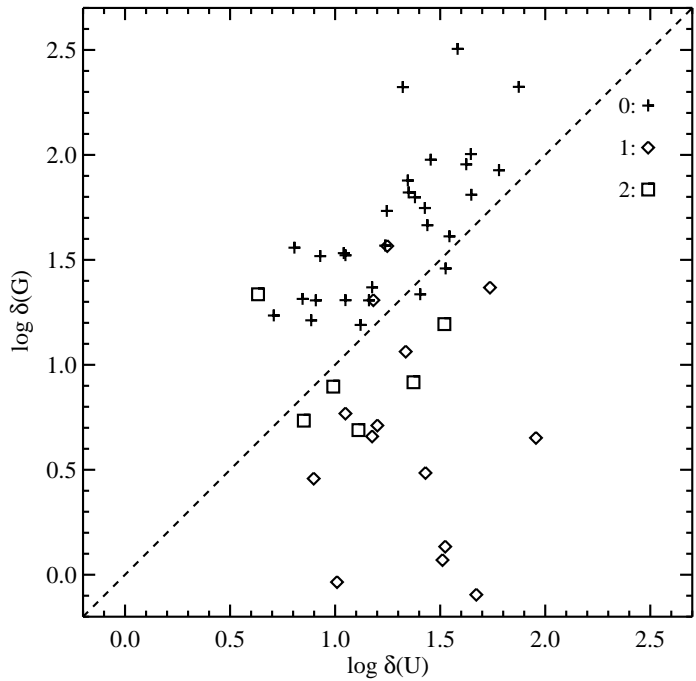}
\vspace{.5cm}
  \caption{Comparison of the O\,{\sc vi} absorber overdensities in 
the photoionization versus hydrostatic equilibrium  cases.}
\label{fig:rhoUG}
\end{figure}

We now compare in figure~\ref{fig:rhoUG} the values of $\delta(U)$ and $\delta(G)$ 
to check the validity of the assumption of hydrostatic equilibrium. For the 
 type 0 absorbers, a Spearman rank correlation test  shows that   
$\delta(G)$ and $\delta(U)$ are correlated at a $>$99\% confidence level.  
 The mean value of their $\delta(G)/\delta(U)$  
overdensity ratio is close to 2.0. It cannot be substantially decreased as 
its dependence on $\Gamma$(H\,{\sc i}), mass fraction in gas and $T$ is small
($\delta(G)/\delta(U) \propto T^{0.17}  \Gamma$(H\,{\sc i})$^{-1/3}$  
$(\Omega_{\rm b}/\Omega_{\rm m})^{-1/3}$). Therefore, this departure
of $\delta(G)/\delta(U)$ from unity may suggest that a fraction
of the observed H\,{\sc i} is not in the O\,{\sc vi} phase. Nevertheless, 
the correlation between $\delta(G)$ and $\delta(U)$ suggests that the 
O\,{\sc vi} type 0 population is roughly in hydrostatic equilibrium.  
For the type 1 absorbers, $\delta(G)$ and $\delta(U)$  are totally uncorrelated
which implies that hydrostatic equilibrium is not a valid assumption for this 
population: low H\,{\sc i} column density absorbers do not trace low density 
regions of the IGM. 
This  further supports that the metal-rich and metal-poor absorbers trace 
different populations.

%\vspace{1.0cm}

%%%%%%%%%%%%%%%%%%%%%%%%%%%%%%%%%%%%%%%%%%%%%%%%%%%%%%%%%%%%%%%%%%%%%%%%%%%

\section{Conclusions and prospectives}\label{sec:concl}

Our large {\it VLT-UVES} sample of 136 O\,{\sc vi} absorbers at 
$\overline{z}=2.28$ towards ten quasars enables a study of the highly 
ionized phase of the IGM, in particular its metal enrichment and contribution
to the cosmic baryon density.

Previous O\,{\sc vi} studies at high $z$ have uncovered a few systems 
with high [O/H] ($>-1.0$) abundances, motivating us to fully investigate this 
class of metal-rich absorbers. Because those systems already identified have 
low H\,{\sc i} column densities (\cite {berg02}; \cite {car02}), our sample 
includes all detected O\,{\sc vi} systems whatever the strength of their 
associated H\,{\sc i} absorption. In contrast, the survey by  Simcoe \etal\ (2004)
only includes systems with N(H\,{\sc i}) $> 10^{13.6}$ cm$^{-2}$. 
We restrict our sample to absorbers with both lines 
of the O\,{\sc vi} doublet clearly detected or, if partially  
blended, enough unambiguous structure to allow for deblending from Lyman lines.
 
Since nearly half of the O\,{\sc vi} absorbers have small line widths, 
$b<12$ km s$^{-1}$ or $T <1.4\times 10^5$ K, photoionization must be the 
dominant ionization process. We thus introduce an observational identification 
criterium to separate the classes of metal-poor (type 0) and metal-rich  
(type 1) absorbers. Selecting a hard UV background flux (see  
\cite {berg02}; \cite {car02} ; \cite {telf02}) and assuming a 0.1 solar 
metallicity yields a column density ratio N(O\,{\sc vi})/N(C\,{\sc iv}) $< 0.25$ 
and $> 0.25$ for the type 0 and type 1 populations, respectively. 

The bulk of the $b$ distributions of these two O\,{\sc vi} populations are
similar except at the highest velocities.  
However, we stress that very few individual components are 
unambiguously broad, a result of blending for complex, multiple systems and 
limited S/N ($\sim 30$-40) in the O\,{\sc vi} range. 

The cosmic oxygen abundance 
is derived under the assumptions of photoionization,  coexistence of O\,{\sc vi}
and C\,{\sc iv} in the same phase, and a solar O/C relative abundance. 
The overall [O/H] 
distribution is clearly bimodal with median values of [O/H] equal to 
$-2.05$ and $-0.33$ for the type 0 and type 1 populations, respectively. 
This is not a consequence of a strong difference in ionization levels  
between the two types of O\,{\sc vi} absorbers.
All of the type 1 O\,{\sc vi} systems and all but two of the type 0 systems 
have  associated C\,{\sc iv} absorption (detection limit of 
N(C\,{\sc iv}) $\approx 1 \times 10^{12}$ cm$^{-2}$).   
%Their log $\big($N(O\,{\sc vi})/N(C\,{\sc iv})$\big)$ distributions, which cover 
%a range of about 2 dex, are similar with mean values both equal to 1.1.
Their N(O\,{\sc vi})/N(C\,{\sc iv}) distributions, which cover 
about two orders of magnitude, are similar with median values both close to 10.   
A high metallicity (median [O/H] $> -1.0$) for the type 1 population is 
still found under different ionization conditions: photoionization together 
with either a gas phase of constant density (overdensity $\delta = 10$ at 
$z = 2.2$) or a temperature fixed by the line width of the main O\,{\sc vi} 
component of each system (always $<14$ km s$^{-1}$). 

The N(N\,{\sc v})/N(O\,{\sc vi}) ratio cannot be used to constrain the 
ionization level of the O\,{\sc vi} phase because the N/O relative abundance
departs from the solar value. In most O\,{\sc vi} absorbers, the 
N\,{\sc v} doublet is weak or undetected and the nitrogen abundance relative
to oxygen is usually well below solar (\cite {berg02}). For a very few 
O\,{\sc vi} systems, the strength of the N\,{\sc v} absorption is similar to 
those of C\,{\sc iv} and  O\,{\sc vi} and the nitrogen abundance may be 
enhanced relative to that of oxygen (\cite {car02}), as also observed in 
quasar associated systems (e.g. \cite {ham00}).  
In our type 1 O\,{\sc vi}  sample,  associated N\,{\sc v} absorption is 
either absent or very weak, except in one case already reported by
Bergeron  \etal\ (2002). This absorber at $z=2.352$ in Q 0329$-$385
was labelled as ``intrinsic'' by these authors because its properties are 
typical of those of associated systems,   
even though it is at 6200 km s$^{-1}$ from the quasar emission redshift.

Our O\,{\sc vi} sample is large enough to derive the first estimate of 
the O\,{\sc vi} column density distribution, although incompleteness becomes   
evident at N(O\,{\sc vi}) $\lesssim 1 \times 10^{13}$ cm$^{-2}$ and 
sample variance may be important at N(O\,{\sc vi}) $\gtrsim 2 \times 10^{14}$ 
cm$^{-2}$. A power-law fit, $f$(N) $\propto$ N$^{-\alpha}$,  yields 
$\alpha \approx 1.7 \pm 0.5$, a value similar to that found for C\,{\sc iv} 
samples, $\alpha$(C\,{\sc iv}) $\simeq 1.8$ (\cite {song01}; \cite {scann05}). 
In contrast, the normalization factor, $f$(N) $ = 2.3\times 10^{-13}$ 
(with an uncertainty of $\sim$ 
30\%) at N(O\,{\sc vi}) = $10^{13.5}$ cm$^{-2}$, is about ten times larger 
than that of C\,{\sc iv} absorbers. We aim to better constrain $f$(N), 
particularly  at large  O\,{\sc vi} column densities, 
by analyzing a larger number of sightlines in a future paper.
There we will also include blended O\,{\sc vi} components associated 
with C\,{\sc iv} absorption.

We use the fit to $f$(N) in a conservative N(O\,{\sc vi}) range, 
$10^{13}$-$10^{15}$  cm$^{-2}$, to estimate the  number density per unit $z$ 
of O\,{\sc vi} absorbers as well as their cosmic density. 
This is a first step 
for correcting $dn/dz$ and  $\Omega_{\rm b}$(O\,{\sc vi}) from incompleteness 
and sample variance.  We find $dn/dz = 74 \pm ^{32} _{8}$. 
Selecting an integration range 
as wide as is usually adopted for C\,{\sc iv} ($10^{12}$-$10^{16}$ cm$^{-2}$) 
would yield a larger value of  $dn/dz$ (and of $\Omega_{\rm b}$(O\,{\sc vi}))
but the uncertainty on the result would then be far too large. 
%(mainly N(O\,{\sc vi})$_{\rm min}$ for $dn/dz$ and N(O\,{\sc vi})$_{\rm max}$ 
%for $\Omega_{\rm b}$(O\,{\sc vi})), but still correct in part the O\,{\sc vi} 
In the case of an unevolving population and a column density lower limit 
equal to that of low $\overline{z} = 0.1$ O\,{\sc vi} surveys, that is 
N(O\,{\sc vi})$_{\rm min} = 10^{13.6}$ cm$^{-2}$ or $w_{\rm r, min}=50$~m\AA\ 
(see e.g. \cite {sem04}), the expected value of $dn/dz$ at $\overline{z} = 2.3$ 
derived from the low $z$  samples is 36 (assuming a 0.1 solar metallicity 
and an ionic fraction O\,{\sc vi}/O = 0.2), whereas that obtained from our 
$f$(N) distribution equals 26. The suggested decline of $dn/dz$ with $z$ is not 
straightforward to interpret as the O\,{\sc vi} absorbers may trace different 
populations at low and high $z$. 

The O\,{\sc vi} cosmic density estimated 
from the individual, observed column densities is 
$\Omega_{\rm b}$(O\,{\sc vi})$\approx 1.5\times 10^{-7}$, i.e. higher than 
previous estimates by a factor 1.3  (\cite {sim04}) and 1.8 (\cite {car02}). 
This increase is due to  the high contribution (35\%) of the type 1 
population to $\Omega_{\rm b}$(O\,{\sc vi}). 
The value derived from the $f$(N) distribution is 2.3 times larger:  
$\Omega_{\rm b}$(O\,{\sc vi})$\approx (3.5\pm ^{3.2}_{0.9}) \times 10^{-7}$.
This illustrates the effects of incompleteness and sample variance in 
our sample, even within the conservative N(O\,{\sc vi}) range adopted. 
To get the element cosmic density, we use the mean 
ionic fraction obtained in the pure photoionization case,  
$\langle$O\,{\sc vi}/O$\rangle$ = 0.15, which yields $\Omega_{\rm b}$(O) 
$\approx (2.3\pm ^{2.1}_{0.6}) \times 10^{-6}$. 
%Under the two other ionization conditions 
%mentioned above, we get similar or higher values of $\Omega_{\rm b}$(O). 

Adopting the solar oxygen abundance given by Anders \& Grevesse (1989), we get
log~$\big(\Omega_{\rm b}$(O)/$\Omega_{\rm b}$(O)$_{\odot}\big) = -2.22$. 
This value is well below that of the metal-rich population and also smaller 
than the metal enrichment of the IGM expected from high $z$ star-forming 
galaxies, 
$\langle$[O/H]$\rangle \sim -1.40$ (\cite {pett99}) or $-1.65$ (\cite {ferr05}).
Although the problem of missing metals at high $z$ (where previously an order 
of magnitude disparity was measured) is now less severe as a  
result of our O\,{\sc vi} survey,  there remains a shortfall of 
observed metals by about a factor of four as compared to those produced by 
star-forming galaxies.  
 
Other properties of the type 1 O\,{\sc vi} absorbers suggest a tight 
link to galactic halos. This population is predominantly 
detected in the vicinity ($\Delta v <$ 450 km s$^{-1}$) of 
 strong H\,{\sc i} systems ($\tau$(Ly-$\alpha$) $>$ 4). This is also the 
case for C\,{\sc iv}-only metal-rich absorbers (O\,{\sc vi} doublet
either outside the observing range, $z<2.0$, or fully blended with saturated 
Lyman lines). In the photoionization case, the type 0 and type 1 O\,{\sc vi}
absorbers have the same  gas overdensity distribution,  
   with a median value 
 $\delta (U) = 22$, but under the assumption of hydrostatic equilibrium
the gas overdensity, $\delta (G)$, distributions of these two populations 
barely overlap. Moreover, the values of $\delta (U)$ and $\delta (G)$ 
are totally uncorrelated for the metal-rich population, whereas they are 
well correlated for the metal-poor population. Consequently, the 
assumption of hydrostatic equilibrium is not valid for the metal-rich 
O\,{\sc vi} population: these absorbers do not trace low density regions 
of the IGM but rather gas outflows in the vicinity of active star-formation 
sites.

If most of the gas in the metal-rich sites is at high temperature  
($T>5 \times 10^5$ K), as  suggested by Pettini (1999), 
oxygen will mainly be in the form of O\,{\sc vii} and O\,{\sc viii} ions  
and their signatures in the very soft X-ray range are not detectable 
with present-day X-ray satellites. For a phase at  lower temperatures,  
$2\times 10^5< T< 5 \times 10^5$~K, the O\,{\sc vi} and H\,{\sc i} species, 
but not C\,{\sc iv} (ionic fraction C\,{\sc iv}/C $< 10^{-2}$), should be 
detectable. We have begun to search for these absorbers with broad 
($b > 50$ km s$^{-1}$), weak Ly-$\alpha$ lines associated with semi-broad,  
weak O\,{\sc vi} doublets ($b > 15$ km s$^{-1}$). This is coupled to  
a statistical analysis of the Ly-$\alpha$ forest in 
simulated spectra (in progress). 

We also plan to acquire deep, multi-band images of the quasar fields 
with several metal-rich O\,{\sc vi} absorbers. If this population does indeed 
trace hot galactic halos, we expect to find a strong correlation with 
star-forming galaxies. Using these images together with spectroscopic 
follow-up of the associated galaxies may help clarify the ejection 
mechanism(s) responsible for the metal-pollution of galactic halos and  
the surrounding IGM.

\begin{acknowledgments}
 S. Herbert-Fort is supported
by the EU under the Marie Curie Early Stage Training programme EARA-EST.
\end{acknowledgments}

\end{document}